\documentclass{article}

\usepackage{authblk}
\usepackage[bookmarks=true]{hyperref}
\usepackage{graphicx,epsfig,color}
\usepackage{cite}
\usepackage[english]{babel}
\usepackage{amssymb,amsfonts,amsmath}
\usepackage{verbatim}
\usepackage{bbm,bbold}
\usepackage{bigints}

  \newcommand{\cF}{{\cal F}}

  \newcommand{\cL}{{\cal L}}
  \newcommand{\cN}{{\cal N}}

\newcommand{\be}{\begin{equation}} \newcommand{\ee}{\end{equation}}
\newcommand{\bea}{\begin{eqnarray}} \newcommand{\eea}{\end{eqnarray}}
\newcommand{\beann}{\begin{eqnarray*}}  \newcommand{\eeann}{\end{eqnarray*}}
\newcommand{\bfig}{\begin{figure}} \newcommand{\efig}{\end{figure}}
\newcommand{\ba}{\begin{array}} \newcommand{\ea}{\end{array}}
\newcommand{\bcen}{\begin{center}} \newcommand{\ecen}{\end{center}}
\newcommand{\btab}{\begin{tabular}} \newcommand{\etab}{\end{tabular}}

\newtheorem{Proposition}{Proposition}[section]

\newtheorem{Theorem}{Theorem}[section]
\newtheorem{Lemma}{Lemma}[section]
\newtheorem{Corrolary}{Corrolary}[section]

\newcommand{\bp}{\begin{Proposition}}   \newcommand{\ep}{\end{Proposition}}
\newcommand{\bt}{\begin{Theorem}}   \newcommand{\et}{\end{Theorem}}
\newcommand{\bl}{\begin{Lemma}}     \newcommand{\el}{\end{Lemma}}
\newcommand{\bc}{\begin{Corrolary}} \newcommand{\ec}{\end{Corrolary}}
\title{Holographic Wilsonian renormalization of a heavy quark moving through a strongly coupled plasma}

\author[1]{Diego Gutiez} 
\author[1]{Carlos Hoyos} 
\affil[1]{\small \em Department of Physics \& Instituto de Ciencias y Tecnolog\'{\i}as Espaciales de Asturias (ICTEA),  \newline Universidad de Oviedo,  c/ Federico Garc\'ia Lorca, 18, 33007, Oviedo, Spain}

\begin{document}

\maketitle

\begin{abstract}
A heavy quark moving through a strongly coupled deconfined plasma has a holographic dual description as a string moving in a black brane geometry. We apply the holographic Wilsonian renormalization method to derive a holographic effective string action dual to the heavy quark. The effective action only depends on the geometry between the black brane horizon and a cutoff localized in the radial direction, corresponding to the IR of the dual theory. We derive RG flow equations for the coefficients in the effective action and show that the force acting on the heavy quark is independent of the position of the cutoff. All the information about the UV is hidden in integration constants of the RG flow equations. This type of approach could be used to improve semi-holographic models where the UV is described by perturbative QCD and the IR by a holographic model. 
\end{abstract}
\newpage

\tableofcontents

%%%%%%%%%%%%%%%%%%%%%%%%%%%%%%%%%%%%%%%%%%%%%%%%%%%%%%
%%%%%%%%%%%%%%%%%%%%%%%%%%%%%%%%%%%%%%%%%%%%%%%%%%%%%%
\section{Introduction}
%%%%%%%%%%%%%%%%%%%%%%%%%%%%%%%%%%%%%%%%%%%%%%%%%%%%%%
%%%%%%%%%%%%%%%%%%%%%%%%%%%%%%%%%%%%%%%%%%%%%%%%%%%%%%

The gauge/gravity duality \cite{Maldacena:1997re,Gubser:1998bc,Witten:1998qj}, aka holography, has been successful in describing some qualitative properties of the quark-gluon plasma, in particular predicting an almost perfect fluid behavior and producing the famous KSS formula for the shear viscosity over entropy density ratio  \cite{Kovtun:2004de,Policastro:2001yc}, that captures the right order of magnitude deduced from hydrodynamic simulations of heavy ion collisions \cite{Luzum:2008cw,vanderSchee:2013pia,Heinz:2013th,Gale:2013da,Niemi:2015qia}.

One of the most indicative signals of the formation of a deconfined quark gluon plasma is jet quenching  (see \cite{Connors:2017ptx} for a comprehensive review). If a highly energetic parton collision takes place close to the surface of the plasma ball, some of the particles produced may escape almost immediately, producing an observable jet, while the path of particles moving in the opposite direction might have to cross a significant portion of the plasma. In this case energy dissipation produced by the interaction with the plasma components weakens or prevents the formation of a back jet. The observation of jet quenching in heavy ion collisions is one of the most convincing evidences in favor of the formation of a deconfined plasma, and one of the most important probes into its properties.

Early on, the duality has been used to model the energy loss of  heavy quarks moving through the plasma \cite{Herzog:2006gh,Gubser:2006bz,CasalderreySolana:2006rq,Chernicoff:2008sa}
%\cite{Hatta:2008tx,Chesler:2008wd,Gubser:2008as,Chesler:2008uy}, 
in order to give an estimation of jet quenching \footnote{Different estimations of jet quenching involve particles moving at the speed of light \cite{Liu:2006ug} or light quarks or gluons \cite{Chesler:2008uy,Gubser:2008as}.}.
The energy loss can be determined from the drag force the quark experiences, which in turn can be obtained from the expectation value of a Wilson line along the quark trajectory.  Following  \cite{Herzog:2006gh,Gubser:2006bz}, the heavy quark maps to a dragging string moving through a black brane geometry in the holographic dual, from which the expectation value of the Wilson line can be extracted.  The string ends at the asymptotic boundary of space and extends all the way to the black brane horizon. According to the usual holographic map, these two regions correspond to energy scales of the UV and IR of the field theory, respectively. This means that the heavy quark motion is sensitive to all the energy scales of the theory, in contrast for instance to hydrodynamic evolution, which is limited to the IR. The sensitivity to multiple energy scales is a challenge for the holographic description. In QCD the gauge coupling becomes weak at high energies, and if this feature was introduced in the holographic model, then the curvature in the dual geometry would become large and stringy corrections to the classical gravity approximation would not be negligible anymore. Introducing stringy corrections may be doable in principle by adding higher derivative terms in the gravitational action. In this fashion, the first corrections away from the strong coupling limit of some properties of the plasma have been studied for conformal theories \cite{Gubser:1998nz, Pawelczyk:1998pb,Buchel:2004di,Buchel:2008ae,Waeber:2015oka,Solana:2018pbk}. However, for a theory with a running coupling like QCD, finding these corrections has not been attempted yet.

As a full string theory description of holographic duals is out of reach at the moment, it is desirable to tackle this issue in a way that avoids dealing with large curvature corrections. A possibility is to adopt a purely phenomenological approach and model the weakly coupled region neglecting possible higher curvature corrections. A second possibility is to follow an effective theory approach and use the holographic model to describe a limited range of energy scales where the theory is strongly coupled. In this category fall the hybrid or semi-holograhic models studied for instance in \cite{Marquet:2009eq,Ficnar:2012yu,Betz:2014cza,Iancu:2014ava,Casalderrey-Solana:2014bpa,Mukhopadhyay:2015smb,Casalderrey-Solana:2015vaa,Casalderrey-Solana:2016jvj}. A drawback of hybrid models is that the theory changes abruptly when the holographic model is used, with no obvious systematic way to improve the method. We can compare this situation with the usual low energy effective field theory approach. In this case the couplings in the effective action are free parameters that have to be fixed by experiments or by matching to the microscopic theory, and a systematic improvement of the low energy effective theory is possible. This is one of the main properties that makes the effective field theory so successful, and hybrid models could be significantly improved if a similar procedure could be implemented for the holographic model. We will take the first steps in this direction by applying  the method of holographic ``Wilsonian'' renormalization \cite{Heemskerk:2010hk,Faulkner:2010jy} to a moving string. Our analysis is an extension of \cite{Casalderrey-Solana:2019vnc}, where we applied this method to static strings in order to extract the quark-antiquark potential.

\begin{figure}[t!]
\begin{center}
\includegraphics[width=10cm]{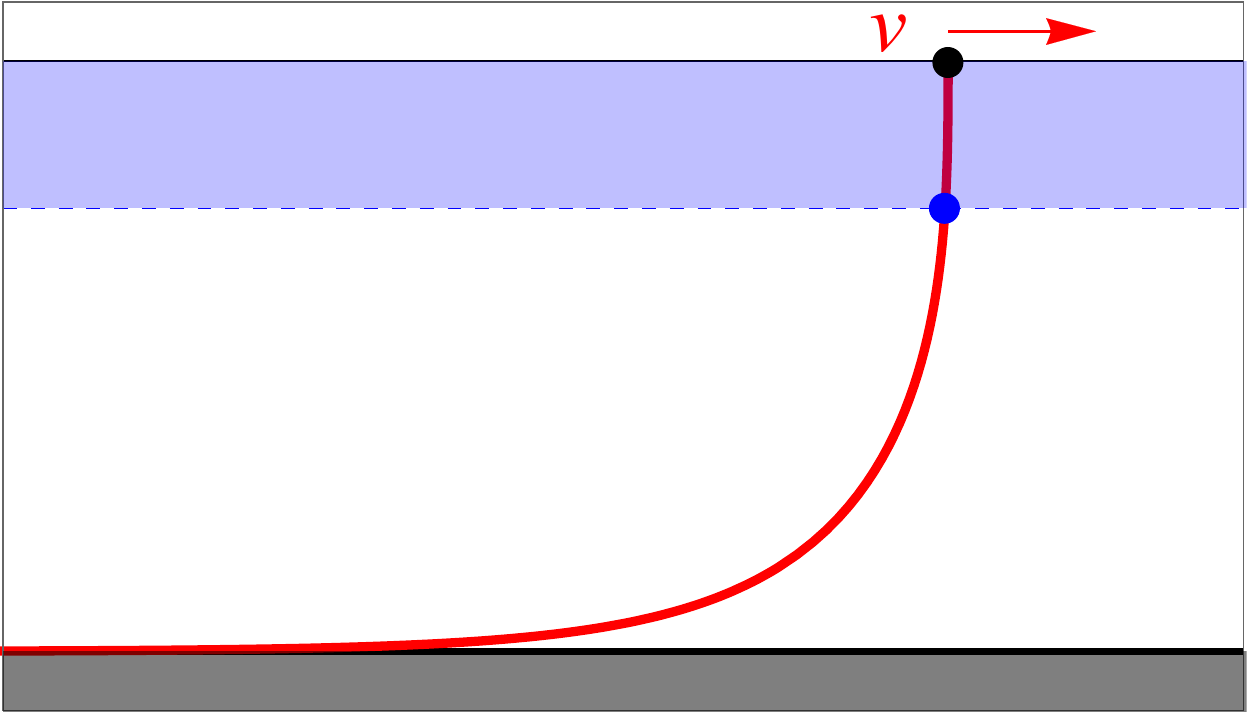}
\caption{\small The holographic dual of a heavy quark moving at speed $v$ is a string (red curve) ending at the asymptotic boundary at the position of the quark (black dot). The strings extends from the asymptotic boundary at the top to the black brane horizon at the bottom of the figure. A cutoff (dashed blue line) is introduced and the shaded region between the boundary and the cutoff is ``integrated out''. One is left with the string in the region between the cutoff and the horizon and determined boundary conditions for the endpoint of the string at the cutoff (blue dot).}\label{fig:dragstring}
\end{center}
\end{figure}

In the Wilsonian method we introduce a cutoff in the dual geometry localized at a fixed distance from the asymptotic boundary. The region between the cutoff and the boundary is replaced by an action for the string endpoints at the cutoff, as sketched in Figure~\ref{fig:dragstring}. The cutoff action ensures that the string satisfies the right boundary conditions, in such a way that the force felt by the heavy quark is independent of the position of the cutoff. The coefficients of the cutoff action satisfy RG flow equations that are determined by the local geometry around the cutoff, in such a way that all the knowledge about the region between the cutoff and the asymptotic boundary is hidden in integration constants of the RG flow equations. Then, for a given IR effective theory, with a holographic dual corresponding to the geometry between the horizon and the cutoff, it is possible to match to multiple UV theories by appropriately tuning the integration constants of the RG flow.  

We will derive the cutoff action and RG flow equations of the coefficients for a heavy quark moving at approximately constant velocity. We will consider a general black brane geometry, but we will also present the results for a IR geometry approximated by an $AdS_5$ black brane, as a simple example to illustrate the method.  We will use these results to compute the drag force, including a couple of contributions proportional to the acceleration of the quark and the jerk, or acceleration rate.  The first can be interpreted as originating from a thermal correction to the mass of the quark, while the second can be thought of as a combination of the Abraham-Lorentz force \cite{Chernicoff:2009re,Chernicoff:2010wg}, due to Larmor radiation, and a viscous contribution produced by the surrounding fluid \cite{Reiten:2019fta}. The two contributions add up giving the total value found in \cite{Banerjee:2015fed}.

The paper is organized as follows. In Sec.~\ref{sec:holoquark} we introduce the holographic description of a heavy quark moving through a strongly coupled plasma as a string in a general black brane geometry. We simplify the analysis by considering a homogeneous and isotropic state for the plasma, and slow variations of the quark trajectory, compared to the inverse temperature. In Sec.~\ref{sec:cutoff} we introduce a cutoff and derive the cutoff action for the string and the RG flow equations for the coefficients. In Sec.~\ref{sec:force}, we apply the general formalism to a specific case where the theory has an IR fixed point and consequently the IR geometry is an $AdS_5$ black brane. We compute the drag force and the corrections proportional to the acceleration and the jerk. In Sec.~\ref{sec:rgflow}, we explain how the cutoff action and RG flow equations can be obtained more generally using the cutoff independence of the string action, and show that they agree with the previous results obtained by direct integration. We end with a discussion of the results in Sec~\ref{sec:discus}. Some technical details of the calculations have been collected in the Appendices. 

%%%%%%%%%%%%%%%%%%%%%%%%%%%%%%%%%%%%%%%%%%%%%%%%%%%%%%
%%%%%%%%%%%%%%%%%%%%%%%%%%%%%%%%%%%%%%%%%%%%%%%%%%%%%%
\section{Holographic description of a heavy quark in a plasma}\label{sec:holoquark}
%%%%%%%%%%%%%%%%%%%%%%%%%%%%%%%%%%%%%%%%%%%%%%%%%%%%%%
%%%%%%%%%%%%%%%%%%%%%%%%%%%%%%%%%%%%%%%%%%%%%%%%%%%%%%

We will start by reviewing the holographic dual to a heavy quark in a high temperature deconfined phase. We will simplify the analysis by imposing that that the heavy quark is moving along one spatial direction in a trajectory of almost constant velocity, with changes in the trajectory that are slow compared to the time scale set by the inverse temperature. This last approximation is necessary in order to apply a low energy effective description. Note that the trajectory is fixed, so we are not considering the dynamics of the quark, only the force of the plasma acting over it.

The holographic dual of the plasma is a five-dimensional geometry with an event horizon extended along four directions, that are identified with the dual field theory directions. A homogeneous and isotropic black brane geometry dual to a strongly coupled $3+1$ plasma can be cast in the general form
\be\label{eq:blackbrane}
ds^2=G_{MN}dx^M dx^N=G_{zz}(z)dz^2+G_{tt}(z)dt^2+G_{xx}(z)\delta_{ij}dx^i dx^j.
\ee
where $(t, x^i)$, $i=1,2,3$ are coordinates along the field theory directions and $z$ is the holographic radial coordinate. The holographic radial direction is identified with energy or length scales in the dual field theory. Low energies, or long wavelengths (IR) correspond to the region close to the horizon, and high energies or short wavelengths (UV) to an asymptotic region far away from the horizon. If the space is asymptotically $AdS_5$, the region approaching the conformal boundary is the far UV. In addition to the metric shown in \eqref{eq:blackbrane}, there can be additional internal directions, but they will not play any role in the following.

We pick coordinates in such a way that there is a horizon at $G_{tt}(z_h)=0$. For the $AdS_5$ black brane metric
\be\label{eq:ads5}
G_{tt}(z)=-\frac{R^2}{z^2} h(z),\ \ G_{zz}(z)=\frac{R^2}{z^2 h(z)},\ \ G_{xx}=\frac{R^2}{z^2}, \ \ h(z)=1-\frac{z^4}{z_h^4}.
\ee
Therefore, $R$ is the $AdS$ radius and the boundary is at $z\to 0$.  The Hawking temperature of the black brane maps to the temperature of the field theory and, as the temperature is increased, the horizon moves towards the asymptotic region, indicating that there are degrees of freedom of higher energy contributing to the plasma. The temperature for the $AdS_5$ black brane geometry is $T=1/(\pi z_h)$. For a general metric we will define the functions
\be\label{eq:fg}
h(z)\equiv \left| \frac{G_{tt}}{G_{zz}}\right|^{1/2},\ \ f(z)\equiv G_{xx}h(z),\ \ g(z)\equiv \frac{G_{xx}}{h(z)}.
\ee
In $AdS_5$ the function $h$ coincides with its usual definition, while $f=|G_{tt}|$ and $g=G_{zz}$. In general, $f$ and $g$ are not equal to metric components, but we will still assume that there is an asymptotic boundary at $z\to 0$ even if the space is not asymptotically $AdS_5$.

We now introduce a heavy quark in the plasma. Considering the mass of the heavy quark to be effectively infinite, the heavy quark maps to a Wilson line and the holographic dual is a string ending at the quark trajectory on the asymptotic boundary. This identification was done originally for static quarks \cite{Maldacena:1998im,Brandhuber:1998er,Rey:1998ik} and later on generalized to quarks in motion  \cite{Herzog:2006gh,Gubser:2006bz,CasalderreySolana:2006rq,Chernicoff:2008sa}. So we do not solve for the quark motion but rather find the force with which the plasma acts when the quark follows a fixed path.

To start with we consider the simplest case of a heavy quark in $\cN=4$ SYM. The dual geometry is $AdS_5\times S^5$, with a string attached to the $AdS_5$ boundary at the location of the Wilson line and localized at a point in the internal space, in this case the $S^5$. To be more precise, this corresponds to a $1/2$ BPS loop that also couples to $\cN=4 $ SYM scalars, but for simplicity we will restrict to this case. We will assume in the following that this setup can be generalized to other holographic duals, i.e. we will use a string to describe a Wilson loop in different geometries, implicitly taking the metric in the string frame and neglecting any motion along internal space directions.

The dynamics of the string are given by the Nambu-Goto action for the embedding functions $X^M$
\be
S_{NG}=-T_s\int d^2\sigma\, \cL_{NG}=-T_s\int d^2\sigma\, \sqrt{-\det g_{ab}(X)},
\ee
where $T_s=1/(2\pi \alpha')$ is the string tension, $\sigma^a$, $a=0,1$ are the worldsheet coordinates and $g_{ab}$ is the induced metric
\be
g_{ab}=G_{MN}(X) \partial_a X^M \partial_b X^N.
\ee

%%%%%%%%%%%%%%%%%%%%%%%%%%%%%%%%%%%%%%%%%%%%%%%%%%%%%%
\subsection{Slowly moving quarks}
%%%%%%%%%%%%%%%%%%%%%%%%%%%%%%%%%%%%%%%%%%%%%%%%%%%%%%

In the first place we will consider a heavy quark that is almost at rest, but can move with a small varying velocity.  Let us assume that the quark moves along one spatial direction. An ansatz for the embedding functions in the static gauge is
\be\label{eq:staticgauge}
t=X^0=\sigma^0, \ X^1=X(t,z),\  X^2=X^3=0,\ z=X^z=\sigma^1.
\ee
The Nambu-Goto action simplifies to
\be\label{eq:SNG}
{\cal L}_{NG}=\sqrt{|G_{tt}| G_{zz}}\sqrt{1+G_{xx}\left(\frac{1}{G_{zz}}(X')^2-\frac{1}{|G_{tt}|}(\dot{X})^2\right)}. 
\ee
Where we have defined the derivatives $X'=\partial_z X$, $\dot{X}=\partial_t X$.

A slowly moving string $\dot{X}\ll 1$ will have a profile that is almost a straight line $X' \ll 1$, as long as we are not too close to the horizon. Then, the string action can be approximated away from the horizon by the quadratic terms
\be\label{eq:L2}
\cL_{NG}\simeq  \sqrt{|G_{tt}| G_{zz}}-\frac{g(z)}{2}(\dot{X})^2+\frac{f(z)}{2}(X')^2\,. 
\ee
Within this approximation the equations of motion take the simple form
\be\label{eq:eomX}
(f X')'-g\ddot{X}=0.
\ee
Assuming that the quark follows a trajectory $x^\mu=(t, x(t),0,0)$, we should fix the position of the string at the boundary to $X(z=0,t)=x(t)$. Since we are considering slow motion we can use a derivative expansion to find the solutions, at least away from the horizon where the function $g(z)$ remains bounded. We expand the profile of the string according to the order in time derivatives
\be
X=X^{(0)}+X^{(1)}+X^{(2)}+\cdots
\ee
Order by order we have the set of equations
\be
\left(f {X^{(0)}}'\right)'=\left(f {X^{(1)}}'\right)'=0, \ \ \left(f {X^{(n)}}'\right)'=g \ddot{X}^{(n-2)}, n\geq 2.
\ee
The equations can be solved recursively. The lowest order solutions are
\be\label{eq:solX0}
X^{(0)}(t,z)=x(t)+p^{(0)}(t) a(z),\ \ X^{(1)}(t,z)=p^{(1)}(t) a(z), \ \ a(z)=\int_0^z \frac{du}{f(u)}.
\ee
where $p^{(0)}(t)$, $p^{(1)}(t)$  are integration constants. The solutions at higher orders take the general form
\be\label{eq:solXn}
X^{(n)}(t,z)= p^{(n)}(t)a(z)+\int_0^z \frac{du}{f(u)}\int_{z_c}^u dv g(v) \ddot{X}^{(n-2)}(t,v).
\ee
We have fixed the limits of the integrals in such a way that
\be\label{eq:cutoffcond}
X(t,z=0)=x(t), \ \ \partial_z X\Big|_{z=z_c}=\frac{1}{f(z_c)} \left( p^{(0)}(t)+p^{(1)}(t)+p^{(2)}(t)+\cdots \right)\equiv \frac{p}{f(z_c)}.
\ee

%%%%%%%%%%%%%%%%%%%%%%%%%%%%%%%%%%%%%%%%%%%%%%%%%%%%%%
\subsection{Fast moving quarks}\label{sec:fastq}
%%%%%%%%%%%%%%%%%%%%%%%%%%%%%%%%%%%%%%%%%%%%%%%%%%%%%%

The previous analysis is a perturbation around a quark at rest. We can generalize it by taking as the unperturbed solution a quark moving at constant velocity. The background profile is the ``dragging string'' found in the original calculations of the drag force  \cite{Herzog:2006gh,Gubser:2006bz}.

The ansatz for the embedding functions is a modification of the ansatz used in the static gauge. Taking $\sigma^0=t$, $\sigma^1=z$,
\be\label{eq:nonstaticgauge}
X^0=t+t_0(z), \ X^1=v t +x_0(z)+X(t,z),\  X^2=X^3=0,\ X^z=z.
\ee
The background profile has constant velocity $v$. The functions $t_0(z)$, $x_0(z)$ determine the profile of the dragging string. $t_0$ determines the gauge and can be conveniently chosen, and $x_0$ is obtained by solving the equations of motion of the background  profile. They are fixed to
\be
t_0'=\frac{G_{xx}}{|G_{tt}|} v x_0',\ \ x_0'=p_0 \sqrt{\frac{|G_{tt}|G_{zz}}{G_{xx}\left(|G_{tt}|G_{xx}-p_0^2 \right)\left(|G_{tt}|-G_{xx}v^2\right)}}
\ee
Where $p_0$ is a constant.  Note that our gauge choice differs from  \cite{Herzog:2006gh,Gubser:2006bz}, where $t_0=0$.

The condition that the solution is real everywhere fixes
\be
p_0^2=|G_{tt}| G_{xx}\Big|_{z=z_*},\ \ v^2=\frac{|G_{tt}|}{G_{xx}}\Big|_{z=z_*}.
\ee
The point $z_*$ is where the speed of the string equals the speed of light on a radial slice, it corresponds to an effective horizon on the string worldsheet outside the black hole horizon. If the geometry is the $AdS_5$ black brane, the solution to these equations is
\be\label{eq:p0v}
z_*= \gamma^{1/4} z_h,\ \ p_0=\frac{R^2}{z_h^2} \gamma v,\ \ \gamma=\frac{1}{\sqrt{1-v^2}}.
\ee

The Nambu-Goto action expanded to quadratic order in the perturbation is
\be\label{eq:SNG2}
\begin{split}
{\cal L}_{NG}\simeq & \sqrt{|G_{tt}| G_{zz} G_{xx}} \sqrt{\frac{|G_{tt}|-G_{xx}v^2}{|G_{tt}|G_{xx}-p_0^2}}+p_0 X'-v \sqrt{\frac{|G_{tt}| G_{xx}^3 G_{zz}}{\left(|G_{tt}|G_{xx}-p_0^2\right) \left(G_{tt}-G_{xx} v^2\right)}}\dot{X}\\
&-\frac{1}{2}g_v(z) (\dot{X})^2+\frac{1}{2}f_v(z) (X')^2.
\end{split}
\ee
Where we have defined the functions
\be\label{eq:fgv}
g_v(z)=(|G_{tt}| G_{xx} G_{zz})^{1/2}\frac{\left(|G_{tt}|G_{xx}-p_0^2\right)^{1/2}}{\left(|G_{tt}|-G_{xx}v^2\right)^{3/2}}, 
	\ \ f_v(z)=(|G_{tt}| G_{xx} G_{zz})^{-1/2} \frac{\left(G_{tt}G_{xx}-p_0^2\right)^{3/2}}{\left(|G_{tt}|-G_{xx} v^2\right)^{1/2}}.
\ee
The terms linear in the perturbation are total derivatives, as expected in an expansion around a solution of the equations of motion. Therefore, they do not affect to the equations of motion of the perturbation. The quadratic terms take the same form in this gauge as those for a slowly moving quark \eqref{eq:L2}, replacing the functions $f, g$ by $f_v, g_v$. Then, we can apply the same type of derivative expansion as for the slowly moving quark and all the results have a straightforward generalization to the case of a fast moving quark.

%%%%%%%%%%%%%%%%%%%%%%%%%%%%%%%%%%%%%%%%%%%%%%%%%%%%%%
%%%%%%%%%%%%%%%%%%%%%%%%%%%%%%%%%%%%%%%%%%%%%%%%%%%%%%
\section{Effective string action with a radial cutoff}\label{sec:cutoff}
%%%%%%%%%%%%%%%%%%%%%%%%%%%%%%%%%%%%%%%%%%%%%%%%%%%%%%
%%%%%%%%%%%%%%%%%%%%%%%%%%%%%%%%%%%%%%%%%%%%%%%%%%%%%%

We now proceed to the derivation of the holographic Wilsonian effective action for the string. First we will use holographic renormalization on the string action to identify the quantity that maps to the force acting on the heavy quark in the dual field theory. For quarks moving along varying trajectories it is not exactly the canonical momentum conjugate to the position of the quark, as it was for quarks moving at constant velocities in \cite{Herzog:2006gh,Gubser:2006bz}, because the canonical momentum depends on the holographic radial coordinate in this case. Next, we will introduce a cutoff localized in the radial direction and ``integrate out'' the region between the asymptotic boundary and the cutoff. As a result, the remaining string action is defined in the region between the horizon and the cutoff, and there is an additional boundary contribution at the cutoff. Finally, we will derive the RG flow equations obtained from varying the position of the cutoff and use them to show that physical observables obtained from the effective action are independent of the cutoff. As the action for fast moving quarks can be mapped straightforwardly to the action for slowly moving quarks, we will restrict the analysis in this section only to the latter, and show results for both cases in the next section.

%%%%%%%%%%%%%%%%%%%%%%%%%%%%%%%%%%%%%%%%%%%%%%%%%%%%%%
\subsection{Force acting on the quark}
%%%%%%%%%%%%%%%%%%%%%%%%%%%%%%%%%%%%%%%%%%%%%%%%%%%%%%

The action \eqref{eq:SNG} determines the string profile for a given quark trajectory $x(t)$ by fixing the position of the string at the asymptotic boundary $z=0$. We will now identify the relevant physical observable, the force acting on the quark. For this it is convenient to momentarily assume that the metric is asymptotically $AdS_5$. Then, when $z\to 0$
\be
f(z)\simeq g(z)\simeq \frac{R^2}{z^2}, 
\ee  
Then, the equation for the profile can be approximated by
\be
X''-\frac{2}{z} X'-\ddot{X}=0.
\ee
Solutions to this equation have a boundary expansion
\be\label{eq:boundexp}
X(t,z)=x(t)-\frac{1}{2}\ddot{x}(t) z^2+\frac{F(t)}{3R^2}z^3+\cdots.
\ee
The string action evaluated on shell is
\be\label{eq:NGos}
\cL_{NG}\simeq  \frac{R^2}{z^2}+\partial_z\left( \frac{f}{2} X' X\right)\,. 
\ee
The action is divergent, we will regularize it introducing a UV cutoff $z_\Lambda$ and adding a boundary counterterm that cancels the divergence when $z_\Lambda \to 0$. The boundary counterterm is determined by the induced metric at the boundary
\be
S_{c.t}=T_s R \int dt \sqrt{-g^b}, \ \ g^b=g^b_{00}=g^b_{00}\Big|_{z_\Lambda}\simeq -\frac{R^2}{z_\Lambda^2}(1-\dot{X}^2(z_\Lambda)).
\ee
Since we are approximating the action to quadratic order,
\be\label{eq:sct}
\sqrt{-g^b}\simeq \frac{R^2}{z_\Lambda}\left(1-\frac{1}{2}\dot{X}(z_\Lambda)^2\right).
\ee
The regularized action is then
\be
S_{\rm string}= \lim_{z_\Lambda \to 0} T_s\int dt \left[ -\int_{z_\Lambda}^{z_h}dz\cL_{NG}+R\sqrt{-g^b}\right]. 
\ee
The string action determines the effective potential felt by the quark $S_{\rm string}=-\int dt V_{q}$, so its variation respect to the position of the quark at the boundary determines the force
\be
\delta S_{\rm string}\simeq  \lim_{z_\Lambda \to 0} T_s\int dt\left[ f X' \delta X\Big|_{z_\Lambda}+\frac{R^2}{z_\Lambda}\ddot{X}(z_\Lambda)\delta X(z_\Lambda)\right]=T_s \int dt F(t) \delta x. 
\ee
Where we have used \eqref{eq:boundexp}. The force acting in the $x$ direction is
\be\label{eq:force}
\cF_x=-\frac{\delta V_{q}}{\delta x}=T_s F(t).
\ee
For a fast moving quark there is a small modification of this result, as the background profile also gives a contribution to the force. The background contribution comes from the linear terms in the action \eqref{eq:SNG2}. Doing a variation, the term with a time derivative drops, and the term with a radial derivative gives the boundary contribution
\be
\delta S_{\rm back}\simeq \lim_{z_\Lambda \to 0} T_s \int dt \,\left(- p_0 \delta X\Big|_{z_\Lambda}\right)=-T_s \int dt\, p_0 \delta x.
\ee
Therefore, the contribution of the unperturbed string profile to the force is
\be\label{eq:forcev}
\cF_x^v=-T_s p_0.
\ee

Let us now find an expression for the slow moving quark, using the solutions \eqref{eq:solX0}, \eqref{eq:solXn}. The derivatives give
\be
f \partial_z X^{(0)}=p^{(0)},\ \ f \partial_z X^{(1)}=p^{(1)}, \ \ f \partial_z X^{(n)}=p^{(n)}+\int_{z_c}^z  g(v) \ddot{X}^{(n-2)}(v).
\ee
Summing over all orders in the expansion, the radial derivative is
\be
f X'=p+\ddot{x}\int_{z_c}^z g(v)  +\ddot{p}\int_{z_c}^z g(v) a(v)+O(\partial_t^4 x, \partial_t^4 p).
\ee
We can do a Taylor expansion if we extract the divergent contribution from the term with a factor $\ddot{x}$:
\be\label{eq:fdX}
f X'=p+\ddot{x}\left[ A(z)-\frac{R^2}{z}\right]+\ddot{p}B(z)+O(\partial_t^4 x, \partial_t^4 p).
\ee
Where we have defined
\be\label{eq:AB}
A(z)= \frac{R^2}{z_c}+\int_{z_c}^z  \left(g(v)-\frac{R^2}{v^2}\right) , B(z)=\int_{z_c}^z  g(v) a(v).
\ee
Then,
\be
X'\simeq -\ddot{x} z +F(t)\frac{z^2}{R^2}+\cdots.
\ee
Where the coefficient that determines the force is
\be\label{eq:Fzm}
F(t)=p+\ddot{x}A(0) +\ddot{p}B(0)+O(\partial_t^4 x, \partial_t^4 p).
\ee
Although it may not look like it at first sight, this expression is independent of $z_c$. Using \eqref{eq:cutoffcond} and the equations of motion \eqref{eq:eomX},
\be
\partial_{z_c} p= \partial_{z_c}(f(z_c) X'(z_c))=g(z_c)\ddot{X}(z_c).
\ee
Then, the derivative of the force is
\be
\partial_{z_c} F=g(z_c)\ddot{X}(z_c)-g(z_c)\ddot{x}-g(z_c)a(z_c)\ddot{p}+O(\partial_t^4 x, \partial_t^4 p)=0+O(\partial_t^4 x, \partial_t^4 p).
\ee
where we have used that $X(z_c)=x+p a(z_c)+O(\partial_t^2 x, \partial_t^2 p)$.

%%%%%%%%%%%%%%%%%%%%%%%%%%%%%%%%%%%%%%%%%%%%%%%%%%%%%%
\subsection{Introducing a cutoff}
%%%%%%%%%%%%%%%%%%%%%%%%%%%%%%%%%%%%%%%%%%%%%%%%%%%%%%

Once we have determined that the force is \eqref{eq:force}, we will formulate a prescription to compute it when there is a cutoff at $z=z_c$ in the geometry such that the string does not reach the $AdS$ boundary, but it is extended between the horizon and the cutoff. 

First, we will split the string action in two parts, corresponding to the integration along the radial coordinate in the UV region $z_c > z>z_\Lambda$ and the IR region $z_h> z > z_c$.
\be
S_{\rm string}=S_{IR}+S_{UV}.
\ee
Where
\be
\begin{split}
&S_{IR}= -T_s\int dt \int_{z_c}^{z_h}dz \cL_{NG},\\
&S_{UV}=  \lim_{z_\Lambda \to 0} T_s\int dt \left[-\int_{z_\Lambda}^{z_c}dz \cL_{NG}+R \sqrt{-g^b}\right]. 
\end{split}
\ee
We will use the on-shell expression \eqref{eq:NGos} and the counterterm \eqref{eq:sct} to evaluate the UV action expanded to quadratic order in the perturbation
\be
S_{UV}\simeq  T_s\int dt \left[\frac{R^2}{z_c}-\frac{f(z_c)}{2} X(z_c) X'(z_c)+\frac{1}{2}\lim_{z_\Lambda\to 0}\left( f(z_\Lambda) X(z_\Lambda) X'(z_\Lambda)+\frac{R^2}{z_\Lambda}\dot{X}(z_\Lambda)^2 \right)\right]. 
\ee
From now on we will denote the position of the string at the cutoff as $X(z_c)\equiv x_c$. As we show in Appendix~\ref{app:suv}, for a slowly moving quark the UV action can be approximated by
\be\label{eq:suv}
S_{UV}\simeq  T_s\int dt \left[M_c-\frac{1}{2}K_c \dot{x}^2-\frac{1}{2 a_c}(x_c-x)^2+\frac{1}{2} m_c(\dot{x}_c- \dot{x})^2-\kappa_c\ddot{x}(x_c-x)+O(\partial_t^4 x, \partial_t^4 x_c)\right]. 
\ee
Where we have defined the coefficients as
\be\label{eq:suvcoef}
M_c=\frac{R^2}{z_c}, \ K_c= A(0), \ a_c=a(z_c),\ \ m_c=\frac{1}{a_c^2}\int_0^{z_c}dv g(v) a(v)^2,\ \ \kappa_c=\frac{1}{a_c}\int_0^{z_c}dv g(v) a(v).
\ee
The string action has now the desired form, it is the Nambu-Goto action in the region of the space between the horizon and the cutoff plus a boundary action defined at the cutoff, that we will use to determine boundary conditions for the string profile in the IR region. Note that the cutoff action depends on both the trajectory of the Wilson loop and the position of the string at the cutoff, and their derivatives. The coefficients in the cutoff action are determined by integrations over the region between the cutoff and the asymptotic boundary, but the dependence on the cutoff is sensitive only to the region close to the radial position of the same.

The total action is stationary respect to changes in the profile of the string that change the position at the cutoff but keep the position at the asymptotic boundary fixed
\be
\delta S_{\rm string}=\delta S_{IR}+\delta S_{UV}=0.
\ee
The variation of the on-shell action in the IR region is
\be
\delta S_{IR} =T_s\int dt fX'\delta X\Big|_{z=z_c}=T_s \int dt \, p \delta x_c.
\ee
The variation of the cutoff action \eqref{eq:suv} is
\be
\delta S_{UV}\simeq  T_s\int dt \left[-\frac{1}{a_c}(x_c-x)-m_c(\ddot{x}_c- \ddot{x})-\kappa_c\ddot{x}+O(\partial_t^4 x, \partial_t^4 x_c)\right]\delta x_c. 
\ee
This results in the boundary condition
\be\label{eq:pRG}
p\simeq \frac{1}{a_c}(x_c-x)+m_c(\ddot{x}_c- \ddot{x})+\kappa_c\ddot{x}+O(\partial_t^4 x, \partial_t^4 x_c),
\ee
which is a mixed boundary condition between the radial derivative $p=f(z_c)X'(z_c)$ and the value of the profile $x_c=X(z_c)$ at the cutoff.

Substituting in the expression for the force \eqref{eq:Fzm} and using $B(0)=-a_c \kappa_c$, $A(0)=K_c$ produces 
\be\label{eq:FRG}
F(t)\simeq  \frac{1}{a_c}(x_c-x)+(m_c-\kappa_c)\ddot{x}_c+(K_c-m_c+2\kappa_c) \ddot{x}+O(\partial_t^4 x, \partial_t^4 x_c).
\ee
We have managed to write the force in terms of the coefficients in the cutoff action, the position of the string at the cutoff and the trajectory of the Wilson line. At this point, all the information about the geometry in the UV region is hidden in the value of the coefficients.

%%%%%%%%%%%%%%%%%%%%%%%%%%%%%%%%%%%%%%%%%%%%%%%%%%%%%%
\subsection{RG flow equations}
%%%%%%%%%%%%%%%%%%%%%%%%%%%%%%%%%%%%%%%%%%%%%%%%%%%%%%

The cutoff action \eqref{eq:suv} could be interpreted as an effective description of the Wilson line after UV degrees of freedom have been integrated out up to the energy scale defined by the cutoff. It has a similar form as a putative Wilsonian action, although it may not be exactly the same as the outcome of an actual field theory calculation. Nevertheless, this line of though can be pursued further in the context of holographic RG flows, in particular we can define RG flow equations for the coefficients in the cutoff action from the dependence on the position of the cutoff in the radial direction. The set of equations we obtain are
\be\label{eq:RGfloweqs}
\begin{split}
& \partial_{z_c} M_c=-\frac{R^2}{z_c^2},\\
& \partial_{z_c} K_c=-g(z_c),\\
&\partial_{z_c} a_c=\frac{1}{f(z_c)},\\
&\partial_{z_c} m_c=-\frac{2}{ f(z_c)}\frac{m_c}{a_c}+g(z_c),\\
&\partial_{z_c} \kappa_c=-\frac{1}{ f(z_c)}\frac{\kappa_c}{a_c}+g(z_c).
\end{split}
\ee
In addition, the position of the string at the cutoff obeys an RG flow equation. Taking into account \eqref{eq:pRG},
\be
\partial_{z_c} x_c=\frac{p}{f(z_c)}=\frac{1}{f(z_c)}\left[ \frac{1}{a_c}(x_c-x)+m_c(\ddot{x}_c- \ddot{x})+\kappa_c\ddot{x}\right]+O(\partial_t^4 x, \partial_t^4 x_c),,
\ee
Using the RG flow equations it is straightforward to show that the force \eqref{eq:FRG} is an RG-flow invariant quantity
\be
\partial_{z_c} F(t)=0+O(\partial_t^4 x, \partial_t^4 x_c).
\ee
Although we used the $AdS$ boundary expansion to help us identify \eqref{eq:force} as the force, in fact we can generalize \eqref{eq:FRG} and also the cutoff action \eqref{eq:suv} to any geometry in the UV region, since it only depends on quantities evaluated at the cutoff and the Wilson line trajectory. Note that RG flow equations determine the coefficients of the cutoff action, that only depend on the local geometry close to the cutoff up to integration constants, where all the information about the UV is hidden. 

%%%%%%%%%%%%%%%%%%%%%%%%%%%%%%%%%%%%%%%%%%%%%%%%%%%%%%
%%%%%%%%%%%%%%%%%%%%%%%%%%%%%%%%%%%%%%%%%%%%%%%%%%%%%%
\section{Force in a heated IR fixed point}\label{sec:force}
%%%%%%%%%%%%%%%%%%%%%%%%%%%%%%%%%%%%%%%%%%%%%%%%%%%%%%
%%%%%%%%%%%%%%%%%%%%%%%%%%%%%%%%%%%%%%%%%%%%%%%%%%%%%%

We will use the results of the previous section to find the first terms that appear in the force when we do a derivative expansion of the quark trajectory in a specific example. Consider a strongly coupled theory that has an IR fixed point. The theory is at finite temperature, but low enough such that the physics is still dominated by the IR conformal theory. The holographic dual for the geometry in the IR region can then be approximated by the $AdS_5$ black brane \eqref{eq:ads5}. The geometry in the UV region is in principle unknown, but all the information about the UV will be hidden in integration constants of the RG flow equations.

%%%%%%%%%%%%%%%%%%%%%%%%%%%%%
\subsection{Profile of the string perturbation in the $AdS_5$ black brane}
%%%%%%%%%%%%%%%%%%%%%%%%%%%%%

The equations of motion for the string profile below the cutoff are \eqref{eq:eomX}. We will use an expansion in plane waves to find solutions:
\be
X(t,z)=\int \frac{d\omega}{2\pi} X_\omega(z) e^{-i\omega t},
\ee
and we will use a similar expansion for the position of the string at the cutoff and the quark trajectory
\be
x_c(t)=\int \frac{d\omega}{2\pi} \widetilde{x}_c(\omega) e^{-i\omega t}, \ \ x(t)=\int \frac{d\omega}{2\pi} \widetilde{x}(\omega) e^{-i\omega t}.
\ee
In addition, in order to remove the explicit dependence on the position of the horizon $z_h$, we do the change of variables
\be
z=z_h u,\ \ w=z_h\omega.
\ee
The position of the cutoff in the new coordinate is $u_c=z_c/z_h$. The equation for the string profile becomes
\be\label{eq:eomXw}
X_\omega''-\left(\frac{2}{u}+\frac{4u^3}{1-u^4}\right)X_\omega'+\frac{w^2}{(1-u^4)^2}X_\omega=0.
\ee
We must impose ingoing boundary conditions at the horizon $X_\omega(u)\sim (1-u^4)^{-i w/4}$ s $u\to 1$. It is possible to do an expansion of the solutions in powers of the frequency, in such a way that they take the form
\be\label{eq:Xwsol}
X_\omega(u)\simeq \widetilde{x}_c\, (1-u^4)^{-i w/4}(1-i w \chi_1(u)-w^2 \chi_2(u)+i w^3 \chi_3(u)+...\cdots).
\ee
The functions $\chi_i$ must be regular at the horizon and we will identify the overall coefficient with the value of the profile solution at the cutoff $X_\omega(u_c)= \widetilde{x}_c$, this fixes the values of the functions $\chi_i$ at the cutoff.  Finally, we have to impose the boundary condition \eqref{eq:pRG}.

%%%%%%%%%%%%%%%%%%%%%%%%%%%%%
\subsection{Cutoff action and boundary conditions}
%%%%%%%%%%%%%%%%%%%%%%%%%%%%%

In the following it will be convenient to define a rescaled version of the metric functions and coefficients of the cutoff action, in terms of the coordinate $u$. First we introduce the rescaled $f$ and $g$ functions
\be\label{eq:rescfg}
\widehat{f}(u)=\frac{1-u^4}{u^2}, \ \ \widehat{g}(u)=\frac{1}{u^2(1-u^4)}.
\ee
And, with these definitions, the rescaled functions that determine the coefficients of the cutoff action
\be\label{eq:rescaledcoef}
\begin{split}
&\widehat{a}(u)=\int_0^u \frac{du_1}{\widehat{f}(u)}, \ \ \widehat{K}(u)=\frac{1}{u}-\int_0^u du_1\left( \hat{g}(u_1)-\frac{1}{u_1^2}\right), \\ 
&\widehat{m}(u)=\frac{1}{\widehat{a}(u_c)^2}\int_0^u du_1\widehat{g}(u_1) \widehat{a}(u_1)^2 ,\ \  \widehat{\kappa}(u)=\frac{1}{\widehat{a}(u_c)}\int_0^u du_1 \widehat{g}(u_1) \widehat{a}(u_1) ,
\end{split}
\ee
As before, we define $\widehat{a}_c=\widehat{a}(u_c)$, $\widehat{K}_c=\widehat{K}(u_c)$, $\widehat{m}_c=\widehat{m}(u_c)$ and $\widehat{\kappa}_c=\widehat{\kappa}(u_c)$. In addition, we introduce $\widehat{f}_c=\widehat{f}(u_c)$ and $\widehat{g}_c=\widehat{g}(u_c)$ to ease notation. The relation to the original functions is
\be
f(z_c)=\frac{R^2}{z_h^2} \widehat{f}_c,\ \ g(z_c)=\frac{R^2}{z_h^2} \widehat{g}_c,\ \ a_c=\frac{z_h^3}{R^2}\widehat{a}_c,\ \ K_c=\frac{R^2}{z_h}\widehat{K}_c,\ \ m_c=\frac{R^2}{z_h} \widehat{m}_c,\ \ \kappa_c=\frac{R^2}{z_h} \widehat{\kappa}_c.
\ee
Using the plane wave expansion and rescaling by the position of the horizon, the boundary condition \eqref{eq:pRG} becomes
\be\label{eq:bcXw}
\widehat{f}_c X_\omega' (u_c)= \left(\frac{1}{\widehat{a}_c}-w^2 \widehat{m}_c\right)\widetilde{x}_c-\left(\frac{1}{\widehat{a}_c}-w^2(\widehat{m}_c-\widehat{\kappa}_c)\right) \widetilde{x}+O(w^4 \widetilde{x}_c,w^4 \widetilde{x}).
\ee
This will determine the value of $\widetilde{x}_c$, that in turn we will use to compute the force \eqref{eq:FRG}.

The details of the calculation of the profile solutions and the derivatives can be found in Appendix~\ref{app:IRsol}. The result, to the order in derivatives we are considering, is
\be\label{eq:xcexp}
\widetilde{x}_c=\widetilde{x}+\sum_{i=1}^3 s_i (-i w)^i\widetilde{x}+O(w^4 \widetilde{x}).
\ee
The coefficients in the expansion are
\be
\begin{split}
& s_1=-\widehat{a}_c,\\
& s_2=\widehat{a}_c \left(\widehat{a}_c-\widehat{\kappa}_c+H_2(u_c)\right),\\
& s_3=\widehat{a}_c\left[\widehat{a}_c(\widehat{m}_c+\widehat{\kappa}_c -\widehat{a}_c)-(2\widehat{a}_c+c_1(u_c)) H_2(u_c)+H_3(u_c)\right].
\end{split}
\ee
The explicit  expression for $c_1$ is given in \eqref{eq:cs} 
\be
c_1(u_c)=-\frac{1}{4}\log(1-u_c^4).
\ee
The definitions of the functions $H_2(u)$ and $H_3(u)$ are in \eqref{eq:Hs}, but we will not use the explicit expressions, rather we compute directly the cutoff values $H_2(u_c)$ and $H_3(u_c)$.

%%%%%%%%%%%%%%%%%%%%%%%%%%%%%
\subsection{Force acting on a slowly moving quark}
%%%%%%%%%%%%%%%%%%%%%%%%%%%%%

Using \eqref{eq:xcexp} to obtain the position of the string at the cutoff $x_c(t)$ and plugging the result in the force \eqref{eq:FRG}, one finds the following terms to leading order in derivatives of the quark trajectory
\be\label{eq:F}
F(t)=\frac{R^2}{z_h^3}\sum_{i=1}^3 F_i\, (z_h \partial_t)^i x+O(\partial_t^4 x).
\ee
With coefficients
\be\label{eq:Fcoef}
\begin{split}
& F_1=\frac{s_1}{\widehat{a}_c}=-1,\\
& F_2=\widehat{K}_c+\widehat{\kappa}_c+\frac{s_2}{\widehat{a}_c}= \widehat{a}_c+\widehat{K}_c+H_2(u_c),\\
& F_3=(\widehat{m}_c-\widehat{\kappa}_c) s_1+\frac{s_3}{\widehat{a}_c}=\widehat{a}_c(2\widehat{\kappa}_c-\widehat{a}_c)-(c_1(u_c)+2\widehat{a}_c) H_2(u_c)+H_3(u_c).
\end{split}
\ee
We can find the explicit values of  $F_2$ and $F_3$ by solving RG flow equations for $\widehat{a}_c$, $\widehat{\kappa}_c$ and $H_2(u_c)$, $H_3(u_c)$. This is done in appendix~\ref{app:RGbh}, the results are
\be\label{eq:RGsolutions}
\begin{split}
&\widehat{a}_c=\frac{1}{4}\log\frac{1+u_c}{1-u_c}-\frac{1}{2}\tan^{-1} u_c=\frac{1}{2} \left(\tanh ^{-1}u_c-\tan ^{-1}u_c\right)+a_{UV},\\
&\widehat{K}_c=\frac{1}{u_c}-\widehat{a}_c+K_{UV},\\
&\widehat{\kappa}_c=-\frac{1}{u_c}+\frac{\widehat{a}_c}{2}+\frac{1}{2\widehat{a}_c}\tanh ^{-1}(u_c^2)+\frac{\kappa_{UV}}{\widehat{a}_c},\\
& H_2(u_c)=-\frac{1}{u_c}+1,\\
& H_3(u_c)=\frac{1}{4}\left(\pi-\log 4\right)-\frac{c_1(u_c)}{u_{c}}+\widehat{a}_c-\frac{1}{2}\tan ^{-1}u_c+\frac{1}{4}\left(2 \log (1+u_c)-3 \log \left(1+u_c^2\right)\right).
\end{split}
\ee
Where $a_{UV}$, $K_{UV}$ and $\kappa_{UV}$ are integration constants that depend on the UV region. In the case where the geometry is just the $AdS_5$ black brane everywhere between the boundary and the horizon, these integration constants vanish $a_{UV}=K_{UV}=\kappa_{UV}=0$.

Plugging the solutions to the RG flow equations \eqref{eq:RGsolutions} in \eqref{eq:Fcoef} produces the cutoff-independent values
\be
F_2=1+K_{UV}, \ \ F_3=\frac{1}{4}\left(\pi-\log 4\right)+2(\kappa_{UV}-a_{UV}).
\ee

Let us now interpret the final result in field theory language. First we multiply by the string tension as in \eqref{eq:force}.  The holographic dictionary maps the $AdS$ radius and the position of the horizon to the 't Hooft coupling $\lambda$ and temperature $T$· of the dual field theory 
\be 
T_s R^2=\frac{R^2}{2\pi \alpha'}= \frac{\sqrt{\lambda}}{2\pi}, \ \ z_h=\frac{1}{\pi T}.
\ee
The force acting on the heavy quark is, to third order in derivatives of the trajectory
\be
\cF_x\simeq \frac{\sqrt{\lambda}}{2\pi} \left( -(\pi T)^2\partial_t x+\pi T \,F_2\, \partial_t^2 x+F_3\, \partial_t^3 x\right)+O(\partial_t^4 x).
\ee
In the first place we observe that the coefficient of the term proportional to the velocity of the quark, $\partial_t x$, agrees with the drag force of \cite{Herzog:2006gh,Gubser:2006bz} and is insensitive to the UV physics, at least in the approximation we are doing of fixing the IR geometry to the $AdS$ black brane solution. 

The coefficient proportional to the acceleration, $\partial_t^2 x$, agrees with the expected thermal correction to the quark mass in pure $AdS$ when $K_{UV}=0$. The quark mass can be determined from the length of a straight string extended between the horizon and a ``flavor brane'' at $z=z_m$
\be
M_q=T_s R^2\int_{z_m}^{z_h} \frac{dz}{z^2}=M_0-\frac{T_s R^2}{z_h}=M_0- \frac{\sqrt{\lambda}T}{2}.
\ee
In the formula above $M_0$ is interpreted as the quark mass at zero temperature. This term modifies the inertial mass of the quark. Indeed if we allowed a very large, but not infinite, mass for the quark, the Newton equation for the quark would be
\be
M_0 \partial_t^2 x=\cF_x.
\ee
Moving the acceleration term in the force to the left side of the equation results in replacing $M_0$ by the thermal corrected mass $M_q$. Therefore, we can interpret $K_{UV}$ as a modification of the thermal mass due to UV physics.

Finally, the coefficient of the jerk or acceleration rate, $\partial_t^3 x$, computed for the $AdS_5$ black brane in \cite{Banerjee:2015fed}, can be interpreted as a combination of the Abraham-Lorentz force produced by the emission of Larmor radiation (see \cite{Chernicoff:2009re,Chernicoff:2010wg}) and a viscous contribution from the surrounding plasma that has been computed in \cite{Reiten:2019fta} following the method developed in \cite{Lekaveckas:2013lha}. In a conformal theory in vacuum the viscous part is absent and the coefficient of the jerk term is $\sqrt{\lambda}/(2\pi)$.  At finite temperature the viscous correction is obtained by subtracting the vacuum contribution from our result. Since the coefficient of the viscous contribution does not depend on temperature, the $T\rightarrow 0$ limit of the acceleration rate contribution does not coincide with the $T=0$ value, as noted in \cite{Banerjee:2015fed}.
%In a conformal theory ($a_{UV}=\kappa_{UV}=0$), the total value is less than half of that in the vacuum, so there is a screening effect even though the coefficient does not depend explicitly on the temperature. 
%Non-zero values of $a_{UV}$ and $\kappa_{UV}$ correspond to changes in the radiation emitted by the quark due to UV physics.

%%%%%%%%%%%%%%%%%%%%%%%%%%%%%
\subsection{Force acting on a fast moving quark}
%%%%%%%%%%%%%%%%%%%%%%%%%%%%%

As we showed in section \ref{sec:fastq}, the quadratic action for the fast moving quark takes the same form as for the slowly moving quark, replacing the functions $f,g$ by the functions $f_v,g_v$  given in \eqref{eq:fgv}. Let us introduce the rescaled coordinates and embedding perturbation
\be\label{eq:scaling}
s=\gamma^{-1/2}\frac{t}{z_h},\ \ u=\gamma^{1/2}\frac{z}{z_h},\ \ Y=\gamma^{3/2}X
\ee
Then, in the $AdS_5$ black brane geometry,  the quadratic terms in string action become
\be
S_{NG}\sim \frac{T_s R^2}{z_h^2}\int ds du \frac{1}{2}\left(\widehat{g} (\partial_s Y)^2-\widehat{f}(\partial_u Y)^2\right). 
\ee
Where we have used the expression for $p_0$ in \eqref{eq:p0v} and $\widehat{f}$, $\widehat{g}$ have the same definition as in \eqref{eq:rescfg}. This allows us to translate directly the results for the slowly moving quark to this case. 

The variation of the action gives a boundary term
\be
\delta S_{NG}\sim \lim_{u\to 0} \frac{T_s R^2}{z_h^2}\int ds\, \frac{1}{u^2}\partial_u Y \delta Y= \lim_{z\to 0} T_s \gamma\int dt \,\frac{R^2}{z^2}\partial_z X \delta X. 
\ee
So we should add a factor of $\gamma$ to the expression we found for the force of the slowly moving quark.

A solution close to the $AdS_5$ boundary has same form as \eqref{eq:boundexp} in the rescaled variables, 
\be
Y\simeq y(s)-\frac{1}{2}\partial_s^2y(s) u^2+\frac{\widehat{F}(s)}{3}u^3+\cdots.
\ee
Then, from \eqref{eq:scaling} and comparing to  \eqref{eq:boundexp}, we have that 
\be
F(t)=\frac{1}{z_h^3}\widehat{F}.
\ee
The coefficient $\widehat{F}$ is the same as \eqref{eq:F} taking into account the rescaling
\be
\widehat{F}(s)=\sum_{i=1}^3 F_i\, (\partial_s)^i y+O(\partial_s^4 y)=\gamma^{3/2}\sum_{i=1}^3 F_i\, (\gamma^{1/2}z_h\partial_t)^i x+O(\partial_t^4 x).
\ee
Then, taking into account the background contribution to the force \eqref{eq:forcev},
\be
\cF_x\simeq \frac{\sqrt{\lambda}}{2\pi} \left(-(\pi T)^2\gamma v -(\pi T)^2\gamma^3\partial_t x+\pi T \,F_2\, \gamma^{7/2}\partial_t^2 x+F_3\, \gamma^4\partial_t^3 x\right)+O(\partial_t^4 x).
\ee
Note that the term proportional to $\partial_t x$ could have been obtained by replacing $v\to v+\partial_t x$ in the first term and expanding to linear order. The $\gamma$ factors appearing in higher derivative terms imply that this expansion requires time derivatives to be much smaller than the temperature for very fast quarks $\partial_t\ll  \gamma^{-1/2}\pi T$.

%%%%%%%%%%%%%%%%%%%%%%%%%%%%%%%%%%%%%%%%%%%%%%%%%%%%%%
%%%%%%%%%%%%%%%%%%%%%%%%%%%%%%%%%%%%%%%%%%%%%%%%%%%%%%
\section{General RG flow equations}\label{sec:rgflow}
%%%%%%%%%%%%%%%%%%%%%%%%%%%%%%%%%%%%%%%%%%%%%%%%%%%%%%
%%%%%%%%%%%%%%%%%%%%%%%%%%%%%%%%%%%%%%%%%%%%%%%%%%%%%%

We have presented an explicit derivation of the cutoff action and RG flow equations from a direct integration of the string action in the UV region using the approximation that changes in the quark trajectory are slow in time compared to the time scale given by the inverse temperature. It is possible to rederive and generalize these results by introducing an ansatz for the cutoff action and using the conditions that the total action should be invariant under changes in the position of the cutoff. In the following we will derive the general RG flow equation for a quark moving in a straight trajectory and compare to the previous results. The interest of this method is that it simplifies somewhat the derivation and allows a systematic extension to non-linear and higher derivative terms, as well as possibly curved trajectories of the quark.

%%%%%%%%%%%%%%%%%%%%%%%%%%%%%%%%%%%%%%%%%%%%%%%%%%%%%%
\subsection{String action and momentum}
%%%%%%%%%%%%%%%%%%%%%%%%%%%%%%%%%%%%%%%%%%%%%%%%%%%%%%

We will work in the static gauge \eqref{eq:staticgauge} and assume that the background metric takes the diagonal form
\be
ds^2=G_{zz}dz^2+G_{tt}dt^2+G_{xx}\delta_{ij}dx^i dx^j.
\ee
We will not make other assumptions about the dependence of the metric components on the coordinates. It will be useful to define a metric for the quark at rest $\gamma_{ab}$, with non-zero components $\gamma_{00}=G_{tt}$ and $\gamma_{11}=G_{zz}$. The induced metric on the string worldsheet is
\be
g_{ab}=\gamma_{ab}+G_{xx}\partial_a X\partial_b X.
\ee
The effective string action consists of the Nambu-Goto action in the IR region of the geometry plus a boundary action defined at the cutoff
\be
S_{\rm string}=S_{IR}+S_c,
\ee
where
\be
S_{IR}=-T_s\int dt \int_{z_c}^{z_h} \cL_{NG}, \ \ S_c=T_s \int dt \cL_c[x_c,\dot{x}_c; z_c].
\ee
The Nambu-Goto Lagrangian density in the case we are considering is given by
\be
\cL_{NG}=\sqrt{-g}=\sqrt{-\gamma}\Delta^{1/2},
\ee
where for later convenience we have defined
\be
\Delta= 1+G_{xx}\gamma^{ab}\partial_a X \partial_b X .
\ee
There is a conserved worldsheet current $\partial_a p_x^a=0$ corresponding to the shift symmetry $X\to X+\delta X$. It can also be identified as the conjugate momentum to the position $X$. It can be obtained from the variation of the string action 
\be
p_x^a=-\frac{\delta \cL_{NG}}{\delta \partial_a X}=-\frac{\sqrt{-\gamma}}{\Delta^{1/2}} G_{xx} \gamma^{ab}\partial_b X.
\ee
Using this expression we can solve for $p_x^0$ and $X'$ in terms of $p_x^1$ and $\dot{X}$.
\be\label{eq:dzX}
p_x^0 =-\Sigma^{1/2} G_{xx} \gamma^{00}\dot{X},\ \ \ X'= -\Sigma^{-1/2} G^{xx} \gamma_{11} p_x^1,
\ee
where we have defined
\be
\Sigma=\frac{|\gamma|-G^{xx} \gamma_{11} (p_x^1)^2}{1+G_{xx}\gamma^{00}(\dot{X})^2}.
\ee

As we have seen, the cutoff action can be obtained from integrating along the radial coordinate the string action in the UV region. The total action should then satisfy the condition that it is stationary under changes of the string profile that preserved the boundary conditions, in particular when the position of the string at the cutoff is displaced keeping the string at the boundary and the horizon fixed
\be
\delta S_{\rm string}=\delta S_{IR}+\delta S_c=0.
\ee
The variation of the IR part is, using the conservation of the momentum $\partial_a p_x^a=0$, 
\be
\delta S_{IR}=T_s \int dt \int_{z_c}^{z_h} dz  \left(p_x^a \partial_a \delta X\right)=T_s \int dt \int_{z_c}^{z_h} dz  \partial_a\left(p_x^a  \delta X\right)=-T_s \int dt \, p_x^1 \delta x_c.
\ee
The variation of the cutoff action is proportional to the Euler-Lagrange equations of $\cL_c$
\be
\delta S_c=T_s \int dt \left[ \frac{\delta \cL_c}{\delta x_c}-\partial_t\left( \frac{\delta \cL_c}{\delta \dot{x}_c}\right)\right]\delta x_c.
\ee
Then, we find the condition
\be\label{eq:bcpx1}
p_x^1\Big|_{z=z_c}=\frac{\delta \cL_c}{\delta x_c}-\partial_t\left(\frac{\delta \cL_c}{\delta  \dot{x}_c}\right)\equiv \delta_{x_c}\cL_c.
\ee

%%%%%%%%%%%%%%%%%%%%%%%%%%%%%%%%%%%%%%%%%%%%%%%%%%%%%%
\subsection{RG flow of the cutoff action}
%%%%%%%%%%%%%%%%%%%%%%%%%%%%%%%%%%%%%%%%%%%%%%%%%%%%%%

The RG flow equations for the cutoff action can be derived from the requirement that the total action should be independent of the position of the cutoff, as it would be the case if we had obtained it by integrating over the UV region. The condition is
\be
\frac{d}{d z_c} S_{\rm string}=\frac{d}{d z_c}S_{IR}+\frac{d}{d z_c}S_c=0.
\ee
The IR term depends on $z_c$ just through the limits of integration
\be
\frac{d}{d z_c}S_{IR}=T_s \int dt \, \cL_{NG}\Big|_{z=z_c}.
\ee
The cutoff action can have an explicit dependence and an implicit dependence in the position of the string at the cutoff
\be
\frac{d}{d z_c}S_c=T_s\int dt\,\left[ \partial_{z_c} \cL_c+(\delta_{x_c} \cL_c)\, \partial_{z_c} x_c\right].
\ee
Note that there is an integration over time, so the RG flow equation for the cutoff action will be defined up to a total derivative 
\be
\partial_{z_c} \cL_c=-(\delta_{x_c} \cL_c)\, \partial_{z_c} x_c-\cL_{NG}\Big|_{z=z_c}+\partial_t V^t.
\ee
From \eqref{eq:dzX} and \eqref{eq:bcpx1} we can derive the RG flow equation for $x_c$
\be
\partial_{z_c} x_c= -\Sigma_c^{-1/2} G^{xx} \gamma_{11}(\delta_{x_c} \cL_c) \Big|_{z=z_c},
\ee
where now
\be
\Sigma_c=\frac{|\gamma|-G^{xx} \gamma_{11} (\delta_{x_c} \cL_c)^2}{1+G_{xx}\gamma^{00}(\dot{x}_c)^2} \Big|_{z=z_c}.
\ee
Using the same formulas, the evaluation of the Nambu-Goto action at the cutoff will be
\be
\cL_{NG}\Big|_{z=z_c}=\sqrt{-\gamma}\Delta_c^{1/2}\Big|_{z=z_c}=|\gamma|\Sigma_c^{-1/2}\Big|_{z=z_c}.
\ee
Adding all the contributions results in the RG flow equation
\be\label{eq:RGLc}
\partial_{z_c} \cL_c=-\sqrt{-\gamma}\left(1+G_{xx}\gamma^{00}(\dot{x}_c)^2 \right)^{1/2}\left(1-G^{xx} \gamma_{11}\frac{(\delta_{x_c} \cL_c)^2}{|\gamma|}\right)^{1/2}\Big|_{z=z_c}+\partial_t V^t.
\ee
This is our final result, it takes the form of a functional equation for the cutoff action $\cL_c$. We do not have a complete solution, but as we will see this equation admits an expansion in derivatives of $x_c$, in such a way that at each order the RG flow equation for the action reduces to RG flow equations for the coefficients in the expansion.

%%%%%%%%%%%%%%%%%%%%%%%%%%%%%%%%%%%%%%%%%%%%%%%%%%%%%%
\subsection{Slowly moving quark}
%%%%%%%%%%%%%%%%%%%%%%%%%%%%%%%%%%%%%%%%%%%%%%%%%%%%%%

We proceed to solve \eqref{eq:RGLc} in the case we have studied before, a slowly moving quark. An obvious ansatz for the cutoff action is to adapt \eqref{eq:suv} to the more general formulas we have derived, in particular the form of the non-derivative term. We will use\footnote{Note that there is invariance under translations in the $x$ direction, so terms that would break this invariance, such as $x^2$ are forbidden.}
\be
\cL_c =\sqrt{-\gamma} \left[\Lambda-\frac{k_0}{2}(x_c-x)^2+\frac{k_1}{2}(\dot{x}_c-\dot{x})^2+\frac{k_2}{2}(\dot{x}_c)^2+\frac{k_3}{2}(\dot{x})^2+O(\partial_t^4 x, \partial_t^4 x_c)\right]. 
\ee
It is implicit in the formula above and the ones that will follow that all the functions depending on the radial coordinate are evaluated at the cutoff. The overall factor is convenient to cancel out similar factors in \eqref{eq:RGLc}. Comparing with \eqref{eq:suv} requires some reshuffling and an integration by parts, the map between the two sets of coefficients is
\be\label{eq:RGmap}
M_c=\sqrt{-\gamma}\Lambda, \ \ \frac{1}{a_c}=\sqrt{-\gamma} k_0,\ \ m_c=\sqrt{-\gamma}(k_1+k_2),\ \ \kappa_c=\sqrt{-\gamma} k_2,\ \ K_c=-\sqrt{-\gamma}(k_2+k_3).
\ee
The variation of the cutoff action with respect to $x_c$ is
\be\label{eq:dxcLc}
\delta_{x_c}\cL_c=-\sqrt{-\gamma}\left( k_0(x_c-x)+(k_1+k_2) \ddot{x}_c-k_1 \ddot{x}\right).
\ee
The derivative with respect to the cutoff position is
\be\label{eq:dzcLc}
\partial_{z_c}\cL_c =\sqrt{-\gamma} \left[\nabla_{z_c}\Lambda-\frac{1}{2}\nabla_{z_c} k_0(x_c-x)^2+\frac{1}{2}\nabla_{z_c} k_1(\dot{x}_c-\dot{x})^2+\frac{1}{2}\nabla_{z_c} k_2(\dot{x}_c)^2+\frac{1}{2}\nabla_{z_c} k_3(\dot{x})^2\right], 
\ee
where we have defined, for any coefficient $C$,
\be
\nabla_{z_c} C=\frac{1}{\sqrt{-\gamma}}\partial_{z_c}\left(\sqrt{-\gamma} C\right)= \left(\partial_{z_c}+\frac{\partial_{z_c}\sqrt{-\gamma}}{\sqrt{-\gamma}}\right) C.
\ee

Introducing \eqref{eq:dxcLc} and \eqref{eq:dzcLc} in \eqref{eq:RGLc}, and expanding to quadratic order in $x$, $x_c$ and derivatives, one finds that terms with derivatives of the cutoff do not completely match with other terms. While terms in \eqref{eq:dzcLc} only involve first time derivatives, terms from \eqref{eq:dxcLc} will include mixed contributions where one factor has two time derivatives and the other none. This is fixed by an appropriate choice of the total time derivative term, in this case all the terms can be matched for 
\be
V^t=-\sqrt{|\gamma|}G^{xx} \gamma_{11}k_0  (x_c-x)\left[\, k_1 (\dot{x}_c-\dot{x})+k_2 \dot{x}_c\right].
\ee
Demanding that the coefficients of terms with different factors of $x$, $x_c$ and their time derivatives vanish independently of each other leads to the RG flow equations for the coefficients:
\be
\begin{split}
&\nabla_{z_c} \Lambda=-1,\\
&\nabla_{z_c} k_0=-G^{xx} \gamma_{11} k_0^2,\\
&\nabla_{z_c} k_1=-G^{xx} \gamma_{11} k_0 (2k_1+k_2),\\
&\nabla_{z_c} k_2=-G_{xx} \gamma^{00}-G^{xx} \gamma_{11} k_0 k_2,\\
&\nabla_{z_c} k_3=G^{xx} \gamma_{11} k_0 k_2.
\end{split}
\ee
If we use the $AdS_5$ black brane solution \eqref{eq:blackbrane} and \eqref{eq:ads5}, the RG flow equations simplify to
\be
\begin{split}
&\partial_{z_c} (\sqrt{-\gamma} \Lambda)=-\frac{R^2}{z_c^2},\\
&\partial_{z_c} (\sqrt{-\gamma}k_0)=-g(z_c) k_0^2,\\
&\partial_{z_c} (\sqrt{-\gamma} k_1)=-g(z_c) k_0 (2k_1+k_2),\\
&\partial_{z_c} (\sqrt{-\gamma} k_2)=g(z_c)(1- k_0 k_2),\\
&\partial_{z_c} (\sqrt{-\gamma} k_3)=g(z_c) k_0 k_2.
\end{split}
\ee
Using the identifications \eqref{eq:RGmap} and the definitions of $f$ and $g$ \eqref{eq:fg}, it is straightforward to recover the RG flow equations \eqref{eq:RGfloweqs}. We thus arrive to the same results both by doing a direct integration of the string action between the boundary and the cutoff and by deriving the RG flow equation for the cutoff action. However, this last method admits in principle a simpler generalization to more complicated cases.

%%%%%%%%%%%%%%%%%%%%%%%%%%%%%%%%%%%%%%%%%%%%%%%%%%%%%%%
%%%%%%%%%%%%%%%%%%%%%%%%%%%%%%%%%%%%%%%%%%%%%%%%%%%%%%%
\section{Discussion}\label{sec:discus}
%%%%%%%%%%%%%%%%%%%%%%%%%%%%%%%%%%%%%%%%%%%%%%%%%%%%%%%
%%%%%%%%%%%%%%%%%%%%%%%%%%%%%%%%%%%%%%%%%%%%%%%%%%%%%%%

The effective IR string action we have derived is valid in the region close to an arbitrary static black brane geometry, assuming homogeneity and isotropy. In principle these conditions could be relaxed, it would be particularly interesting to study time-dependent geometries emulating the dual to a heavy ion collision, see section VII of \cite{Berges:2020fwq} for a recent review on the topic. An example of this type is the calculation of the drag force in \cite{Chesler:2013cqa}, for a plasma formed by the collision of two infinite sheets with finite energy density in a conformal theory. Another natural extension would be to use the general method presented in Sec.~\ref{sec:rgflow} for less constrained quark trajectories, allowing sudden changes in the trajectory and motion in more than one spatial direction. The general method could also be used to compute nonlinear contributions of acceleration to energy and momentum loss, that in vacuum show in Li\`enard's formula and the Abraham-Lorenz force \cite{Mikhailov:2003er,Chernicoff:2009re}.

One should keep in mind that we are making an assumption by taking the Nambu-Goto action for the string. In many cases the holographic dual is not presented as a ten-dimensional geometry, but has been truncated to five dimensions, and the metric is presented in the Einstein frame. The string action will then be modified by some additional factors. Similarly, if the model is bottom-up, with not known string theory construction behind, then the string action dual to a Wilson line may be chosen in a different way. Nevertheless, in all these cases the method we have presented here can be easily generalized, the derivative expansion of the action will be similar even if the detailed dependence of the coefficients on the geometry can change.

The holographic Wilsonian renormalization method applied here to the string action can be used more generally, for other observables like Wilson lines in different representations or 't Hooft lines as well as for observables obtained from the background geometry, such as the expectation value of local operators. A fully effective description would involve introducing the cutoff and deriving the RG flow equations for the holographic actions dual to all the observables under consideration. It would be interesting to combine the holographic Wilsonian approach with other phenomenological approaches trying to fit QCD lattice data or experiments. Among these, we have the traditional holographic QCD models where the gravitational action is adjusted \cite{Erlich:2005qh,DaRold:2005mxj,Gursoy:2007cb,Gursoy:2007er,Jarvinen:2011qe} or, more recently, the application of machine learning \cite{Hashimoto:2018ftp,Hashimoto:2018bnb,Akutagawa:2020yeo,Hashimoto:2020jug} and Monte Carlo techniques \cite{Jokela:2020auu} to constrain the background geometry. In both cases, the holographic description of UV physics is expected to be problematic due to the asymptotic freedom of QCD. The holographic Wilsonian formalism limits the range of energy scales where the model is applied, so it could be used to avoid this issue without introducing additional assumptions.

%%%%%%%%%%%%%%%%%%%%%%%%%%%%%%%%%%%%%%%%%%%%%%%%%%%%%%%
%%%%%%%%%%%%%%%%%%%%%%%%%%%%%%%%%%%%%%%%%%%%%%%%%%%%%%%
\section*{Acknowledgments}
%%%%%%%%%%%%%%%%%%%%%%%%%%%%%%%%%%%%%%%%%%%%%%%%%%%%%%%
%%%%%%%%%%%%%%%%%%%%%%%%%%%%%%%%%%%%%%%%%%%%%%%%%%%%%%%

We thank Jorge Casalderrey-Solana and Niko Jokela for useful comments. D.G. has been partially supported by Oviedo University through the program Plan de Apoyo y Promoci\'on de la Investigaci\'on  PAPI-19-PF-09 and by the Principado de Asturias through a Severo Ochoa fellowship PA-20-PF-BP19-044. D.G. and C.H. have been partially supported by the Spanish grant PGC2018-096894-B-100 and by the Principado de Asturias through the grant GRUPIN-IDI/2018 /000174.

\appendix

%%%%%%%%%%%%%%%%%%%%%%%%%%%%%%%%%%%%%%%%%%%%%%%%%%%%%%%
%%%%%%%%%%%%%%%%%%%%%%%%%%%%%%%%%%%%%%%%%%%%%%%%%%%%%%
\section{Derivation of the cutoff action}\label{app:suv}
%%%%%%%%%%%%%%%%%%%%%%%%%%%%%%%%%%%%%%%%%%%%%%%%%%%%%%%
%%%%%%%%%%%%%%%%%%%%%%%%%%%%%%%%%%%%%%%%%%%%%%%%%%%%%%

In this Appendix we explain how to derive the cutoff action from direct integration of the string action between the boundary and the cutoff. We can expand the solutions using the derivative expansion. Their form is \eqref{eq:solX0}, \eqref{eq:solXn}, and they satisfy the conditions \eqref{eq:cutoffcond}. This gives the following simplifications, 
\be
f(z_c) X'(z_c)= p, \ \ \lim_{z_\Lambda\to 0} X(z_\Lambda)=x,\ \ \lim_{z_\Lambda\to 0} \dot{X}(z_\Lambda)=\dot{x}.
\ee
It remains to evaluate the derivative at the boundary. From \eqref{eq:fdX}
\be
\lim_{z_\Lambda\to 0}\left( f(z_\Lambda) X(z_\Lambda) X'(z_\Lambda)+\frac{R^2}{z_\Lambda}\dot{X}(z_\Lambda)^2 \right)= x F(t)+\lim_{z_\Lambda\to 0}\frac{R}{z_\Lambda}\left(\ddot{x} x+\dot{x}^2 \right).
\ee
The last term is a total derivative and we can drop it, while $F(t)$ is given by the expression in \eqref{eq:Fzm}. We are left with
\be
S_{UV}\simeq  T_s\int dt \left[\frac{R^2}{z_c}+\frac{1}{2} (F\, x-p\,  x_c)\right]. 
\ee
Instead of $p$, we would like the action to depend on the position of the string at the cutoff and the boundary $x_c$, $x$ and on their derivatives. In order to solve for $p$, first we integrate \eqref{eq:fdX} between the boundary and the cutoff
\be
x_c-x=p a(z_c)+\ddot{x} C(z_c) +\ddot{p} D(z_c)+O(\partial_t^4 x, \partial_t^4 p).
\ee
Where we have defined
\be
C(z)=\int_0^z\frac{dv}{f(v)}\left( A(v)-\frac{R^2}{v}\right),\ \ D(z)=\int_0^z dv \frac{B(v)}{f(v)}.
\ee
We can further simplify these expressions using the explicit form of $A$ and $B$ \eqref{eq:AB}, the definition of $a(z)$ in \eqref{eq:solX0} and integration by parts
\be
\begin{split}\label{eq:CD}
&C(z)=\int_0^zdv a'(v)\int_{z_c}^v du\,g(u)=a(z)\int_{z_c}^z du g(u)-\int_0^z dv g(v) a(v),\\
&D(z)=\int_0^z dv a'(v) \int_{z_c}^v du\, g(u)a(u)=a(z) B(z)-\int_0^z dv g(v) a(v)^2.
\end{split}
\ee
Evaluating at the cutoff we obtain
\be
C(z_c)=-\int_0^{z_c} dv g(v) a(v)=B(0),\ \  D(z_c)=-\int_0^{z_c} dv g(v) a(v)^2.
\ee
Then, we can solve for $p$ as
\be
p=\frac{1}{a(z_c)} (x_c-x)-\frac{C(z_c)}{a(z_c)}\ddot{x}-\frac{D(z_c)}{a(z_c)^2} (\ddot{x}_c-\ddot{x})+O(\partial_t^4 x, \partial_t^4 x_c).
\ee
Introducing this in \eqref{eq:Fzm} to the same order, 
\be
F=\frac{1}{a(z_c)} (x_c-x)+\frac{a(z_c) C(z_c)-D(z_c)}{a(z_c)^2} (\ddot{x}_c-\ddot{x})+\left(A(0)-\frac{C(z_c)}{a(z_c)}\right)\ddot{x}+O(\partial_t^4 x, \partial_t^4 x_c).
 \ee
Finally, the UV action can be arranged, up to a total derivative in time, to be \eqref{eq:suv} with coefficients 
\be
M_c=\frac{R^2}{z_c},\  K_c=A(0), \ a_c=a(z_c),\ \ m_c=-\frac{D(z_c)}{a(z_c)^2},\ \ \kappa_c=-\frac{C(z_c)}{a(z_c)}.
\ee
Using that 
\be
C(z_c)=-\int_0^{z_c} dv g(v) a(v), \ \ D(z_c)=-\int_0^{z_c} dv g(v) a(v)^2,
\ee
leads to the expressions in \eqref{eq:suvcoef}.

%%%%%%%%%%%%%%%%%%%%%%%%%%%%%%%%%%%%%%%%%%%%%%%%%%%%%%%
%%%%%%%%%%%%%%%%%%%%%%%%%%%%%%%%%%%%%%%%%%%%%%%%%%%%%%
\section{Solution for the string profile in the $AdS$ black brane}\label{app:IRsol}
%%%%%%%%%%%%%%%%%%%%%%%%%%%%%%%%%%%%%%%%%%%%%%%%%%%%%%%
%%%%%%%%%%%%%%%%%%%%%%%%%%%%%%%%%%%%%%%%%%%%%%%%%%%%%%

In this Appendix we give explicit formulas for the the solutions of the perturbation of the string profile in the $AdS_5$ black brane geometry.
Introducing \eqref{eq:Xwsol} into the equation \eqref{eq:eomXw} and expanding in $w$, we get the following equations at each order in the expansion
\be
\chi_i''+\frac{\widehat{f}'}{\widehat{f}}\chi_i'-\frac{1}{\widehat{f}} j_i=0,
\ee
where 
\be
j_1=1,\ \ j_2=\sigma(u)+2u \chi'_1+\chi_1,\ \ j_3=\sigma(u) \chi_1+2u \chi'_2+\chi_2,
\ee
and  we have defined
\be
\sigma(u)=\frac{1}{u^2}+\frac{u^2}{1+u^2}.
\ee
The general form of the solutions that satisfy regularity at the horizon $u=1$ is 
\be
\chi_i(u)=c_i(u_c)+\int_{u_c}^u du_1\frac{J_i(u_1)}{\widehat{f}(u_1)}, \ \ J_i(u)=\int_1^u du_2 j_i(u_2).
\ee
where $c_i(u_c)$ are integration constants. 

The expansion of \eqref{eq:Xwsol} in powers of $w$ leads to
\be\label{eq:Xwsol2}
X_\omega(u)\simeq \tilde{x}_c \left(1-iw\Gamma_1(u)-w^2\Gamma_2(u)+iw^3 \Gamma_3(u)+\cdots\right),
\ee
where we have introduced the functions
\be\label{eq:gamma}
\begin{split}
&\Gamma_1(u)=\chi_1(u)+\frac{1}{4}\log(1-u^4),\\
& \Gamma_2(u)=\chi_2(u)-\frac{1}{32}\log^2(1-u^4)+\frac{1}{4}\Gamma_1(u)\log(1-u^4),\\
&\Gamma_3(u)=\chi_3(u)+\frac{1}{4}\Gamma_2(u) \log(1-u^4)-\frac{1}{32}\Gamma_1(u) \log^2(1-u^4)+\frac{1}{384}\log(1-u^4)^3.
\end{split}
\ee
The boundary condition at the cutoff is $\Gamma_i(u_c)=0$, this fixes the integration constants of the solutions to
\be\label{eq:cs}
c_1(u_c)=-\frac{1}{4}\log(1-u_c^4),\ \ c_2(u_c)=\frac{1}{32}\log^2(1-u_c^4),\ \ c_3(u_c)=-\frac{1}{384}\log^3(1-u_c^4).
\ee

In order to determine $\widetilde{x}_c$ we need to compute $X_\omega'$ given by \eqref{eq:Xwsol2} at the cutoff, introduce it in the boundary condition \eqref{eq:bcXw} together with  \eqref{eq:xcexp} and solve order by order in $w$. The result is
\be
\begin{split}
& s_1=\widehat{a}_c \widehat{f}_c \Gamma_1'(u_c),\\
& s_2=\widehat{a}_c \left(\widehat{f}_c \Gamma_2'(u_c)+\widehat{a}_c (\widehat{f}_c \Gamma_1'(u_c))^2-\widehat{\kappa}_c )\right),\\
& s_3=\widehat{a}_c\left[\widehat{f}_c \Gamma_3'(u_c)+\widehat{a}_c^2  (\widehat{f}_c \Gamma_1'(u_c))^3+\widehat{a}_c \widehat{f}_c \Gamma_1'(u_c)\left(2 \widehat{f}_c\Gamma_2'(u_c) -\widehat{m}_c-\widehat{\kappa}_c \right)\right].
\end{split}
\ee
These expressions can be further simplified. First, from the definitions \eqref{eq:gamma} one can derive the following relations
\be
\begin{split}
&\Gamma_1'(u_c)=\frac{J_1(u_c)}{\widehat{f}_c}-c_1'(u_c),\\
&\Gamma_2'(u_c)=\frac{J_2(u_c)}{\widehat{f}_c}-c_1(u_c) \Gamma_1'(u_c)-c_2'(u_c),\\
&\Gamma_3'(u_c)=\frac{J_3(u_c)}{\widehat{f}_c}-c_1(u_c) \Gamma_2'(u_c)-c_2(u_c) \Gamma_1'(u_c)-c_3'(u_c).
\end{split}
\ee
In addition, one can extract some constant factors
\be\label{eq:Js}
J_2(u)=c_1(u_c) J_1(u)+H_2(u),\ \ J_3(u)=c_2(u_c)J_1(u)+H_3(u),
\ee
where
\be\label{eq:Hs}
\begin{split}
H_2(u)=\int_1^u du_1\left( \sigma(u_1)+2c_1'(u_1) J_1(u_1)+\int_{u_c}^{u_1}du_2\, \frac{J_1(u_2)}{\widehat{f}(u_2)}\right),\\
H_3(u)=\int_1^u du_1\left(\chi_1(u_1)\sigma(u_1)+2c_1'(u_1) J_2(u_1)+\int_{u_c}^{u_1}du_2\, \frac{J_2(u_2)}{\widehat{f}(u_2)}\right).
\end{split}
\ee

%%%%%%%%%%%%%%%%%%%%%%%%%%%%%%%%%%%%%%%%%%%%%%%%%%%%%%%
%%%%%%%%%%%%%%%%%%%%%%%%%%%%%%%%%%%%%%%%%%%%%%%%%%%%%%
\section{RG flow in $AdS_5$ black brane}\label{app:RGbh}
%%%%%%%%%%%%%%%%%%%%%%%%%%%%%%%%%%%%%%%%%%%%%%%%%%%%%%%
%%%%%%%%%%%%%%%%%%%%%%%%%%%%%%%%%%%%%%%%%%%%%%%%%%%%%%

In this Appendix we explain how to obtain the solutions to the RG flow equations in the case where the IR geometry is approximately an $AdS_5$ black brane.
The RG flow equations for the rescaled coefficients of the cutoff action \eqref{eq:rescaledcoef} $\widehat{a}_c$, $\widehat{K}_c$ and $\widehat{\kappa}_c$ are
\be
\partial_{u_c} \widehat{a}_c=\frac{1}{\widehat{f}_c},\ \ \partial_{u_c} \widehat{K}_c=-\widehat{g}_c,\ \ \partial_{u_c} \widehat{\kappa}_c=-\frac{1}{\widehat{a}_c \widehat{f}_c}\widehat{\kappa}_c+\widehat{g}_c.
\ee
From \eqref{eq:rescfg},  $\widehat{f}_c=(1-u_c^4)/u_c^2$ and $\widehat{g}_c=1/(u_c^2(1-u_c^4)$. In pure $AdS_5$ the coefficients satisfy the conditions 
\be
\widehat{a}_c\Big|_{u_c}=0,\ \ \widehat{a}_c\widehat{\kappa}_c\Big|_{u_c=0}=0.
\ee
Otherwise there will be integration constants that depend on the UV geometry in a non-trivial way.

By direct integration of the equations, one finds the following values for the coefficients
\be
\begin{split}
&\widehat{a}_c=\frac{1}{4}\log\frac{1+u_c}{1-u_c}-\frac{1}{2}\tan^{-1} u_c=\frac{1}{2} \left(\tanh ^{-1}u_c-\tan ^{-1}u_c\right)+a_{UV},\\
&\widehat{K}_c=\frac{1}{u_c}-\widehat{a}_c+K_{UV},\\
&\widehat{\kappa}_c=-\frac{1}{u_c}+\frac{\widehat{a}_c}{2}+\frac{1}{2\widehat{a}_c}\tanh ^{-1}(u_c^2)+\frac{\kappa_{UV}}{\widehat{a}_c}.
\end{split}
\ee
We will now derive RG flow equations for $H_2(u_c)$ and $H_3(u_c)$ obtained from evaluating \eqref{eq:Hs} at the cutoff. We will be using that $J_1(u_c)=u_c-1$, the relations
\be
c_1'(u)=\frac{u}{f(u)},\ \  c_2'(u)=c_1(u) c_1'(u),\ \   c_3'(u)=c_2(u)  c_1'(u),
\ee
and, from \eqref{eq:Js},
\be
\partial_{u_c} J_2(u)=c_1'(u_c) J_1(u)-\frac{J_1(u_c) J_1(u)}{\widehat{f}_c}=\frac{J_1(u)}{\widehat{f}_c}.
\ee
Then, we derive the following RG flow equations
\be
\partial_{u_c}H_2(u_c)=\sigma(u_c)+2c_1'(u_c) J_1(u_c)-\frac{J_1(u_c)^2}{\widehat{f}_c}=\sigma(u_c)+2\frac{J_1(u_c)}{\widehat{f}_c}+\frac{J_1(u_c)^2}{\widehat{f}_c}.
\ee
\be
\begin{split}
&\partial_{u_c}H_3(u_c)=c_1(u_c) \sigma(u_c)+2c_1'(u_c) J_2(u_c)-\frac{J_2(u_c) J_1(u_c)}{\widehat{f}_c}\\
&+\int_1^{u_c} du \left(\underset{1/\widehat{f}_c}{\underbrace{\left[c_1'(u_c)-\frac{J_1(u_c)}{\widehat{f}_c}\right]}}\sigma(u)+2c_1'(u) \frac{J_1(u)}{\widehat{f}_c}
+\frac{1}{\widehat{f}_c}\int_{u_c}^u du_1\,\frac{J_1(u_1)}{f(u_1)}\right)\\
&=c_1(u_c) \sigma(u_c)+2\frac{J_2(u_c)}{\widehat{f}_c}+\frac{J_2(u_c) J_1(u_c)}{\widehat{f}_c}\\
&+\frac{1}{\widehat{f}_c}\underset{H_2(u_c)}{\underbrace{\int_1^{u_c} du \left(\sigma(u)+ 2c_1'(u) J_1(u)
+\int_{u_c}^u du_1\,\frac{J_1(u_1)}{f(u_1)}\right)}}\\
&=c_1(u_c) \sigma(u_c)+2\frac{J_2(u_c)}{\widehat{f}_c}+\frac{J_2(u_c) J_1(u_c)}{\widehat{f}_c}+\frac{H_2(u_c)}{\widehat{f}_c}.
\end{split}
\ee
Finally, with the explicit expressions for $\sigma$, $J_1$ and $\widehat{f}$, we get the simple RG flow equation for $H_2$:
\be
 \partial_{u_c} H_2(u_c)=\frac{1}{u_c^2}.
\ee
Similarly, if in addition we use that $J_2(u_c)=c_1(u_c)J_1(u_c)+H_2(u_c)$, the RG flow equation for $H_3$ takes the simpler form
\be
 \partial_{u_c} H_3(u_c)=\frac{c_1(u_c)}{u_c^2}+\frac{H_2(u_c)(J_1(u_c)+3)}{\widehat{f}_c}.
\ee
We can integrate both equations taking into account that $H_2(1)=0$, $H_3(1)=0$, the solutions are
\be
\begin{split}
 & H_2(u_c)=-\frac{1}{u_c}+1,\\
& H_3(u_c)=\frac{1}{4}\left(\pi-\log 4\right)-\frac{c_1(u_c)}{u_{c}}+\widehat{a}_c-\frac{1}{2}\tan ^{-1}u_c+\frac{1}{4}\left(2 \log (1+u_c)-3 \log \left(1+u_c^2\right)\right).
\end{split}
\ee
Note that $H_2$, $H_3$ and their boundary conditions are defined in the IR region of the geometry, so there are no additional integration constants associated to the RG flow equations of $H_2(u_c)$ and $H_3(u_c)$.

\bibliographystyle{JHEP}

\bibliography{biblio}

\end{document}